\documentclass[aps,pr,twocolumn,
superscriptaddress,groupedaddress,nofootinbib,floatfix]{revtex4-1}  
\usepackage{soul}
\usepackage{amsfonts}
\usepackage{amssymb}
\usepackage{amsmath}
\usepackage{amsthm}
\usepackage{graphicx}
\usepackage{indentfirst}
\usepackage{appendix}
\usepackage{bbold}
\usepackage[small,bf]{caption}
\usepackage{slashed}
\usepackage{tensor}
\usepackage[caption=false]{subfig}
\usepackage{dsfont}
\usepackage{enumitem}
\usepackage{xcolor}
\usepackage[pdftitle={bfg_gribov},colorlinks=true]{hyperref}

\usepackage{tikz}
\usepackage[compat=1.1.0]{tikz-feynman}

\usepackage[normalem]{ulem}
\newcommand\redsout{\bgroup\markoverwith{\textcolor{red}{\rule[0.4ex]{3pt}{0.7pt}}}\ULon}

\providecommand{\U}[1]{\protect\rule{.1in}{.1in}}

\newcommand{\p}{\partial}

\newcommand{\e}{\ensuremath{\mathrm{e}}}

\usepackage{color}
\usepackage{listings}
\definecolor{darkgreen}{rgb}{0,0.35,0}
\definecolor{Rood}{rgb}{1, 0, 0}

\begin{document}

\title{\textbf{Gauge copies and the fate of background independence in Yang-Mills theories:\\ a leading order analysis}}

\author{O.~C.~Junqueira}
\email{octavio@if.uff.br}
\affiliation{UFF $-$ Universidade Federal Fluminense, Instituto de F\'{\i}sica, Campus da Praia Vermelha, Av.~Litor\^anea s/n, 24210-346, Niter\'oi, RJ, Brasil.}
\affiliation{UFRJ $-$ Universidade Federal do Rio de Janeiro, Instituto de F\'{\i}sica, Caixa Postal 68528, 21941-972, Rio de Janeiro, RJ, Brasil.}

\author{I.~F.~Justo}
\email{ijusto@id.uff.br}
\affiliation{UFF $-$ Universidade Federal Fluminense, Instituto de F\'{\i}sica, Campus da Praia Vermelha, Av.~Litor\^anea s/n, 24210-346, Niter\'oi, RJ, Brasil.}

\author{D.~S.~Montes}
\email{douglasmontes@id.uff.br}
\affiliation{UFF $-$ Universidade Federal Fluminense, Instituto de F\'{\i}sica, Campus da Praia Vermelha, Av.~Litor\^anea s/n, 24210-346, Niter\'oi, RJ, Brasil.}

\author{A.~D.~Pereira}
\email{adpjunior@id.uff.br}
\affiliation{UFF $-$ Universidade Federal Fluminense, Instituto de F\'{\i}sica, Campus da Praia Vermelha, Av.~Litor\^anea s/n, 24210-346, Niter\'oi, RJ, Brasil.}

\author{R.~F.~Sobreiro}
\email{rodrigo\_sobreiro@id.uff.br}
\affiliation{UFF $-$ Universidade Federal Fluminense, Instituto de F\'{\i}sica, Campus da Praia Vermelha, Av.~Litor\^anea s/n, 24210-346, Niter\'oi, RJ, Brasil.}

\begin{abstract}
In this work we investigate the effects of the Gribov prescription to get rid of zero-modes of the Faddeev-Popov operator, at one-loop order in perturbation theory, in the Landau-DeWitt gauge. {Quantum fluctuations are taken around a transverse background gauge field.} The one-loop effective action is explicitly computed, and the behavior of the gauge and ghost fields propagators are carefully investigated. At one-loop and for generic transverse background configurations the effective action is found to be \textit{not} background invariant, as expected, due to {a non-vanishing background contribution}. The gauge field propagator has the same form as in the case {where the} background {is
a trivial} field, {\it i.e.} with complex {conjugate} poles, which are modified by the corresponding gap equation. The ghost-anti-ghost propagator still displays its enhanced $\sim p^{-4}$ behavior.
\end{abstract}

\maketitle

\section{Introduction}

The description of all elementary forces of nature with the exception of gravity is encoded in the quantum-field-theoretic setting of the Standard Model of Particle Physics. The tiles of the particle physics mosaic are the Yang-Mills theories. Such theories display a remarkable property of being asymptotically free at high energies and perturbation theory is a very effective computational tool. Conversely, at low energies, Yang-Mills theories become non-perturbative and relying on perturbative calculations turns out to be  impossible due to the appearance of an infrared (IR) Landau pole, signaling the breakdown of perturbation theory. Nevertheless, many different frameworks to deal with the {strongly}-correlated regime of Yang-Mills theories were developed in the last decades and much progress has been achieved \cite{Brambilla:2014jmp}. Concretely, lattice methods \cite{Maas:2011se}, effective models \cite{Tissier:2010ts,Reinosa:2017qtf,Oxman:2018dzp}, holographic dualities \cite{Witten:1998zw,Dudal:2018ztm}, the reformulation of the gauge-fixing procedure at the non-perturbative level \cite{Vandersickel:2012tz}, functional methods \cite{Binosi:2009qm,Cyrol:2016tym,Huber:2018ned}, are different frameworks that aim at grasping properties of Yang-Mills theories in the IR regime. Successfully, the interplay of such approaches have brought key new insights in the past years and is paving the way for a clearer understanding behind, e.g., color confinement.

The path integral formulation {of} gauge theories, when treated in a continuum space(time), often requires the introduction of a gauge-fixing term in order to remove the {overcounting} of gauge equivalent field configurations. The celebrated method to achieve a proper gauge-fixing of the path integral in consistency with perturbative unitarity is the so-called Faddeev-Popov (FP) method \cite{Faddeev:1967fc}.  Such a procedure involves the assumption that there is a choice of gauge condition that selects one gauge field representative per gauge orbit, i.e., if a given gauge field satisfies the gauge condition then all the other gauge field configurations that belong to the same gauge orbit do not satisfy the same gauge constraint. Next to that, the gauge condition is inserted {in} the path integral at the price of {introducing} the FP determinant. Another assumption behind the FP method is that such a determinant is well-defined in the sense that zero-modes are absent\footnote{More precisely, the standard Faddeev-Popov procedure assumes the positivity of the Faddeev-Popov operator.}. For practical purposes, we often choose gauge conditions that contain differential operators and, with very {few} exceptions, that are Lorentz invariant. The problem is: in his seminal work, Gribov showed that the famous Landau gauge condition actually does not satisfy all the assumptions underlying the FP method \cite{Gribov:1977wm}. In the same year, Singer proved that this is not a particular pathology of the Landau gauge but a feature of the non-trivial geometric structure of Yang-Mills theories \cite{Singer:1978dk}. Explicitly, such gauges are not ideal in the sense that more than one gauge field representative in a given gauge orbit satisfy the gauge fixing condition. The existence of such spurious field configuration after gauge fixing is what is referred to as the Gribov problem and such configurations are known as Gribov copies. Reviews on the topic can be found in \cite{Vandersickel:2012tz,Sobreiro:2005ec,Pereira:2016inn}. 

Nevertheless, in perturbation theory, i.e., for small fluctuations around the trivial vacuum $A^{a}_\mu=0$ (with $A^a_\mu$ being the gauge field), the assumptions of the FP method are met. However, at the non-perturbative level, those assumptions are not satisfied anymore suggesting that Gribov copies can play a role non-perturbatively and therefore should be taken into account in order to capture the correct IR behavior of Yang-Mills theories. In \cite{Gribov:1977wm}, Gribov proposed to restrict the path integral domain to a region which is free of a large set of copies; the infinitesimal ones. Such copies are connected to the original configuration through an infinitesimal gauge transformation. This region, known as the Gribov region, is defined by the set of transverse field configurations which lead to a positive FP operator in the Landau gauge. It features remarkable geometric properties: it is bounded in every direction; it is convex; and all gauge orbits cross it at least once, ensuring that such a restriction does not remove any physical content of the theory \cite{DellAntonio:1991mms}. However, the theory is not completely free of copies due to the presence of those generated by large gauge transformations, see \cite{vanBaal:1991zw}. Ideally, one would restrict the path integral to the so-called ``Fundamental Modular Region", which is by definition free of all Gribov copies. In practice, however, it is only known how to implement the restriction of the path integral to the Gribov region, so far. The restriction is imposed, formally, by demanding that the FP ghosts two-point function does not develop any non-trivial pole. This can be implemented order by order in the coupling constant. In \cite{Gribov:1977wm}, this was implemented at leading order, and in \cite{Zwanziger:1989mf} Zwanziger proposed an all order approach to implement the restriction by means of a different method. The equivalence of {those} approaches, i.e., the one obtained by Gribov's prescription up to all orders and the one derived by Zwanziger, was established in \cite{Capri:2012wx}. We will not revisit the original construction in the Landau gauge since the reader can find it in details in \cite{Vandersickel:2012tz}.

Although the main {idea} of this (partial) solution of the Gribov problem could be extended to different gauges, technical challenges show up. In particular, the FP operator in the Landau gauge is Hermitian, a feature which ensures the reality of its eigenvalues and, therefore, an order relation between them. Hence, it is meaningful to look for a region where the FP operator is ``positive". Nonetheless, this is not a general property and a simple departure from the Landau gauge to linear covariant gauges (LCG) -- where the longitudinal part of the gauge field is fixed but non-vanishing -- already spoils this feature. A prominent exception is the maximal Abelian gauge (MAG) which features a Hermitian FP operator and allows for an explicit construction of a Gribov-region analogue, see \cite{Capri:2005tj,Capri:2006cz,Capri:2008ak,Capri:2010an}.

Effectively, the restriction to the Gribov region is implemented by the introduction of a non-local term known as the horizon function. Such a non-locality can be cast in a local form through the introduction of localizing fields. The resulting action is the so-called Gribov-Zwanziger (GZ) action \cite{Zwanziger:1989mf}. It is local, renormalizable at all orders in perturbation theory and effectively implements the restriction to the Gribov region \cite{Zwanziger:1989mf,Vandersickel:2012tz}. Next to the horizon function, a mass parameter known as the Gribov parameter naturally emerges. Such a parameter is consistently fixed by a gap equation and therefore it is not a new free parameter. We refer the reader not familiar with all those details to \cite{Vandersickel:2012tz}. More recently, it was observed that the theory defined by the GZ action suffers from infrared instabilities and favor the formation of lower-dimensional condensates, see \cite{Dudal:2007cw,Dudal:2008sp,Dudal:2011gd}. They are associated to the gluon field as well as to the localizing auxiliary fields. In order to take into account the existence of such condensates in the starting point theory, the refined Gribov-Zwanziger (RGZ) action was proposed \cite{Dudal:2008sp}. Such {a} refined theory features a gluon propagator which is finite at zero momentum {at tree level}. In fact, such a propagator fits well very recent {gauge-fixed} lattice data \cite{Cucchieri:2007rg,Bogolubsky:2009dc,Cucchieri:2010xr,Cucchieri:2016jwg} for the gluon propagator and puts the RGZ action as a suitable candidate description of the strongly-correlated regime of Yang-Mills theories. Recently, the one-loop corrections to the ghost-gluon vertex was computed in the soft gluon configuration and, again, the results qualitatively agree with lattice simulations, see \cite{Mintz:2017qri} and \cite{Gracey:2012wf}. 

The (R)GZ action has an intriguing property: it explicitly breaks the BRST invariance in a soft way. This means that the break vanishes in the deep ultraviolet (UV) regime, restoring the well-known properties of the BRST-invariant FP quantization. Such a breaking {of the BRST symmetry} was intensively studied in the past decade \cite{Dudal:2009xh,Sorella:2009vt,Baulieu:2008fy,Capri:2010hb,Dudal:2012sb,Dudal:2014rxa,Pereira:2013aza,Pereira:2014apa,Capri:2014bsa,Tissier:2010ts,Serreau:2012cg,Serreau:2015yna,Lavrov:2013boa,Moshin:2015gsa,Schaden:2014bea,Cucchieri:2014via}. Later on, it was shown that the RGZ action can be expressed in a BRST-invariant form provided that a gauge-invariant dressed field is introduced together with a Stueckelberg-like field \cite{Capri:2015ixa,Capri:2015nzw,Capri:2016aqq,Capri:2016gut,Capri:2017abz,Capri:2017bfd,Capri:2018ijg,Mintz:2018hhx}. Hence, it is possible to reconcile the restriction {of the path integral} to the Gribov region with BRST invariance. This allowed for a ``non-perturbative" BRST quantization rule which implements the restriction to a region free of a large set of Gribov copies in LCG and Curci-Ferrari gauges, \cite{Capri:2017bfd,Pereira:2016fpn}, giving {a} universal meaning to the horizon function \cite{Capri:2018ijg}. Moreover, the BRST-invariant construction was extended to the MAG as well, \cite{Capri:2015pfa}. For LCG, it was proved that the local and BRST-invariant action is renormalizable at all orders in perturbation theory, \cite{Capri:2017bfd}. In addition, BRST-invariance ensures the independence of correlation function of BRST-closed operators from gauge parameters, \cite{Capri:2016gut}, a non-trivial issue in the BRST non-invariant setting. Hence, the state-of-the-art of the quantization of Yang-Mills theories by restricting {the configuration space of the path integral} to the Gribov region in covariant gauges is summarized by a local, renormalizable and BRST-invariant setup. 

In practical calculations using functional techniques and/or at finite-temperature, c.f., \cite{Aguilar:2008xm,Braun:2007bx,Braun:2010cy,Reinosa:2014ooa} the use of the background field method (BFM) is very frequent, \cite{Abbott:1981ke,Weinberg:1996kr}. Such a method consists in splitting the gauge field in a fixed background part and a fluctuation piece. Its power consists in retaining explicit gauge invariance at the level of the effective action. However, the splitting is an artificial tool and observables should not depend on the choice of the background, i.e., the theory must be background independent. To illustrate such a fact, in the Yang-Mills action, the background $\bar{A}^a_\mu$ and fluctuation $a^a_\mu$ fields always appear as a sum $A^{a}_\mu = \bar{A}^a_\mu + a^a_\mu$. In the FP gauge-fixing procedure, background and fluctuation fields are not treated as a sum, but such split enters as a BRST-exact term. Then, a natural question arises at this point: How does Gribov copies manifest themselves in the BFM and what are their effects to background independence? The study of Gribov copies in the background Landau gauge, the so-called Landau-DeWitt (LDW) gauge, was {performed} in recent papers, see \cite{Canfora:2015yia,Canfora:2016ngn,Dudal:2017jfw,Kroff:2018ncl}. The existence of infinitesimal Gribov copies is conditioned to the existence of zero-modes of the FP operator which depends on the choice of background. Hence, it is perfectly conceivable that the (partial) resolution of the Gribov problem for a given background will bring background-dependent information to the new effective action. In other words, the resolution will potentially spoils background gauge invariance. Such a fact was identified and dealt with in \cite{Dudal:2017jfw}. In fact, as it is known, background invariance is related to (anti-)BRST symmetry (\textit{c.f.} \cite{Binosi:2013cea}) and the authors of \cite{Dudal:2017jfw} made use of the BRST-invariant reformulation of the refined Gribov-Zwanziger action, \cite{Capri:2017bfd}, to propose an action that preserves background gauge invariance while takes into account the existence of infinitesimal Gribov copies in the Landau-DeWitt gauge. A different path was followed in \cite{Kroff:2018ncl}. In all these works, the authors have explicitly chosen a constant background configuration, although their main outcome is not restricted to that particular case. However, the GZ action in the Landau-DeWitt gauges was not actually derived in the sense of the standard Gribov-Zwanziger action but rather a proposal was made and different consistency checks were made. The main goal of this work is to derive the leading order effective action, while taking into account the existence of infinitesimal Gribov copies in the Landau-DeWitt gauge, for a generic transverse background $\bar{A}^a_\mu$, by following Gribov's original prescription. Along the derivation, we make explicit several subtle points about the imposition of the no-pole condition in this case (some of them, very similar to those found out in the case of the maximal Abelian gauge, \cite{Capri:2005tj}). Not surprisingly, we find that the vacuum energy depends on the choice of the background, just in agreement with \cite{Canfora:2015yia,Canfora:2016ngn,Dudal:2017jfw,Kroff:2018ncl}. Such a dependence, however, vanishes when the background is chosen to be constant. We also discuss how the restriction to the Gribov region in the Landau-DeWitt gauge affects the propagators of the gluon and ghost fields. This work can be viewed as a stepping stone towards the derivation of the Gribov-Zwanziger action in the Landau-DeWitt gauge, which should, naturally, be connected/equivalent to the results of \cite{Dudal:2017jfw} due to the requirement of background independence. Moreover, the explicit derivation of the no-pole condition can be seen as an inspiration to the case of quantum gravity where a Gribov problem is expected to exist and the use of the background field method is ubiquitous.

The paper is organized as follows: in Sect.~\ref{gpldw} we set up our conventions and state the Gribov problem in the Landau-DeWitt gauge; in Sect.~\ref{nopolegapeq}, the no-pole condition is worked out at leading order and the gap equation which fixes the Gribov parameter is derived. Sect.~\ref{2pointgreensfunct} is devoted to {a discussion about} the gluon and ghost propagators in the presence of the Gribov restriction. At the end of Sect.~\ref{2pointgreensfunct} the constant background configuration is investigated as a first exercise.
Conclusions and an outlook is presented afterwards. Important but lengthy derivations and Feynman rules are collected in {the} appendices.

\section{The Gribov problem in the Landau-DeWitt gauge}
\label{gpldw}

In this section we {set up} definitions, conventions and {notations} employed to describe Yang-Mills theories in the Landau-DeWitt gauge. The Faddeev-Popov quantization is performed and the Gribov problem in this gauge is stated.

\subsection{Preliminaries}

The basic ingredients {of} Yang-Mills {theories} is a local gauge symmetry defined by a semi-simple Lie group $G$ and a gauge field. For our purposes, we consider the $SU(N)$ group\footnote{At some point we will consider the particular case $N=2$.}. The scenario of the model is a four-dimensional {flat} Euclidean space-time. The Lie algebra-valued gauge field is represented by $A_\mu^a$, where Greek indices refer to space-time indices running through $\{0,1,2,3\}$, and lower case Latin indices are associated to the adjoint-representation of the gauge group and {run} through $\{1,2,\ldots,N^2-1\}$. Throughout this paper the notation of integrals will be simplified according to
\begin{equation*}
\int \mathrm{d}^{4}x ~\to~ \int_{x}
\quad \text{and}\quad
\int \frac{\mathrm{d}^{4}k}{(2\pi)^{4}} ~\to~ \int_{k^{4}}\,,
\end{equation*}
and
\begin{equation*}
\int \frac{\mathrm{d}^{d}k}{(2\pi)^{d}} ~\to~ \int_{k^{d}}
\,.
\end{equation*}
The standard (explicit) notation will be used whenever ambiguity or confusion is possible.

The Yang-Mills action describing the dynamics of the gauge field is
\begin{eqnarray}
S_{\mathrm{YM}} ~=~ 
\frac{1}{4}
\int_{x} ~F^a_{\mu\nu} F^a_{\mu \nu}
\,,
\label{ym1}
\end{eqnarray}
with
$
F^a_{\mu\nu} ~=~ 
\p_{\mu}A^a_{\nu} - \p_{\nu}A^a_{\mu} +gf^{abc}A^b_{\mu}A^c_{\nu}\,\label{fstr1}
$
being the field strength. The parameter $g$ stands for the coupling constant, and $f^{abc}$ represents the structure constants of the gauge group. The action \eqref{ym1} is invariant under infinitesimal gauge transformations\footnote{The action \eqref{ym1} is also invariant under large gauge transformations. Nevertheless, large gauge transformations will not be considered in this work.} of the form
\begin{equation}
\delta A_\mu^a=-D_\mu^{ab}\xi^b\,,\label{gt1}
\end{equation}
with $\xi^a$ being the infinitesimal gauge parameter and the covariant derivative in {the} adjoint representation is defined as $D_\mu^{ab}=\delta^{ab}\p_\mu-gf^{abc}A_\mu^c$.

In order to employ the BFM and implement the LDW condition, the gauge field is decomposed in a classical background $\bar{A}_\mu$ and a fluctuation $a_\mu$, i.e.,
\begin{eqnarray}
A^a_\mu=\bar{A}^a_\mu + a^a_\mu\,.\label{decomp1}
\end{eqnarray}
The field strength {$F^a_{\mu\nu}$} decomposes as
\begin{equation}
    F^a_{\mu\nu}=\bar{F}^a_{\mu\nu} + \bar{D}^{ab}_\mu a^b_\nu - \bar{D}^{ab}_\nu a_\mu^b + gf^{abc} a^{b}_{\mu}a^{c}_{\nu}\,,\label{fstr2}
\end{equation}
where $\bar{D}^{ab}_{\mu} =
\delta^{ab}  \p_{\mu}  - gf^{abc}\bar{A}^{c}_{\mu}$ is the covariant derivative with respect to the {background field} $\bar{A}_{\mu}$ and $\bar{F}^a_{\mu\nu} = \p_\mu\bar{A}^a_\nu - \p_\nu\bar{A}^a_{\mu} +gf^{abc}\bar{A}^b_\mu\bar{A}^c_\nu$ is the background field strength. Within decomposition \eqref{decomp1}, the gauge transformations \eqref{gt1} become
\begin{eqnarray}
    \delta\bar{A}_\mu^a &=&0 \,,
    \nonumber\\
    \delta a_\mu^a &=& -D^{ab}_\mu\xi^b\,,\label{gt2}
\end{eqnarray}
with $D^{ab}_\mu=\delta^{ab}\p_\mu-gf^{abc}( \bar{A}^c_\mu+a^c_\mu)$.

The LDW gauge fixing amounts to impose the constraint
\begin{equation}
\bar{D}_{\mu}^{ab} a^{b}_{\mu} = 0\,,\label{ldwgaugecond} 
\end{equation}
to the fluctuation field while maintaining covariance with respect to the background. At {the} quantum level, this condition can be implemented by the FP quantization method which is straightforward and leads to the gauge fixed partition function
\begin{equation}
    Z=\mathcal{N}\int[da]\,\mathrm{det}(\mathcal{M}^{ab})\,\exp\left(-S_{\mathrm{YM}}-S_{\mathrm{gf}}\right)
    \,,
    \label{z1}
\end{equation}
with $\mathcal{N}$ being a normalization constant, $S_{gf}$ the gauge fixing action given by
\begin{equation}
    S_{\mathrm{gf}}=
    -\int_{x} \frac{\left( \bar{D}_{\mu}a_{\mu} \right)^{2}}{2\alpha}
    \,,
    \label{gf1}
\end{equation}
and with
\begin{equation}
{\cal M}^{ab}=-\bar{D}_\mu^{ac} D_\mu^{cb}
\label{fpop}
\end{equation}
standing for the FP operator. An important feature of the FP operator is that it is Hermitian (See App.~\ref{Ap1}). Such a procedure relies on the positivity condition of the FP operator $\mathcal{M}^{ab}$.

Within the standard FP gauge fixing procedure, the functional determinant of $\mathcal{M}^{ab}$, in equation \eqref{z1}, can be rewritten in a functional integral form to obtain a contribution to the Boltzmann factor, by means of the introduction of a pair of anti-commuting ghosts, $(c^{a},\, \bar{c}^{a})$, known as FP ghosts:
\begin{equation}
\det \mathcal{M}^{ab} ~=~ \int [d\bar{c}][dc]\, \e^{-S_{gh}}
\end{equation}
with
\begin{equation}
\label{ghaction}
S_{gh} ~=~ 
- \int_{x} ~ \bar{c}^{a}(x) \mathcal{M}^{ab} c^{b}(x)
\,.
\end{equation}

In order to effectively implement the LDW gauge condition, \eqref{ldwgaugecond}, the limit $\alpha \to 0$ must be taken. After gauge fixing, the gauge symmetry \eqref{gt2} is no longer valid. Nevertheless, background gauge transformations of the form
\begin{eqnarray}
    \delta_B\bar{A}_\mu^a&=&\bar{D}^{ab}_\mu\omega^b\,,\nonumber\\
    \delta_B(\mathrm{other\;fields})^a&=&gf^{abc}(\mathrm{other\;fields})^b\omega^c 
    \,,\label{gt3}
\end{eqnarray}
are still manifest, \cite{Weinberg:1996kr}.

\subsection{The Gribov problem}
\label{thegribovproblemsubsection}

 Besides assuming the positivity of the FP operator, it is assumed that there is only one field configuration satisfying the gauge condition \eqref{ldwgaugecond}. In Gribov's seminal work \cite{Gribov:1977wm} it was shown that this is not true. Indeed, two gauge configurations connected by an infinitesimal gauge transformation,
\begin{equation}
    \widetilde{a}_\mu^a=a_\mu^a - D_\mu^{ab}\xi^b\;.\label{gt4}
\end{equation}
i.e., two configurations belonging to the same gauge orbit. By applying the operator $\bar{D}_\mu$ in both sides and imposing the LDW gauge condition to both configurations one easily finds 
\begin{equation}
   \mathcal{M}^{ab}\xi^b=0\;.\label{gceq1}
\end{equation}
Equation \eqref{gceq1} is called Gribov (infinitesimal) copies equation and $\widetilde{a}_\mu^a$ is a Gribov copy of the field $a_\mu^a$. A nontrivial {and normalizable} solution of the Gribov copies equation implies that the FP procedure is not complete. Moreover, equation \eqref{gceq1} {states} that infinitesimal copies of a configuration $a_\mu^a$ exist {as} the {normalizable} zero modes of the FP operator. Thence, the functional integral \eqref{z1} still {considers} several spurious configurations after the implementation of the FP gauge-fixing procedure. 

To eliminate them (at least at infinitesimal level) the prescription of \cite{Gribov:1977wm} will be followed. To do so, the fact that $\mathcal{M}$ is Hermitian is crucial: the equation \eqref{gceq1} can be treated as an eigenvalue equation for the FP operator. The eigenvalues will be necessarily real and thus obey an order relation. For each configuration $a_\mu^a$, an eigenvalue spectrum will be defined. Whenever the spectrum has a zero mode, a region of infinitesimal copies is identified. In order to eliminate these infinitesimal Gribov copies we will impose the restriction to the region $\Omega$ where the FP operator is positive definite:
\begin{align}
\label{omegadefinition}
\Omega = \left\{
    a_{\mu}; \ \bar{D}_{\mu}a_{\mu} =0; \
    \mathcal{M}^{ab} > 0
    \right\}
    \,.
\end{align}
The standard procedure is to implement a restriction on the functional integral domain up to the Gribov horizon in order to eliminate infinitesimal Gribov copies. In the Landau gauge, this is motivated by two facts: first, that for a given configuration close enough to the horizon its corresponding gauge copy lies on the other side of that horizon \cite{Gribov:1977wm}; and second, that every gauge orbit intersects the horizon at least once \cite{DellAntonio:1991mms}. Hence, every configuration outside the first Gribov region is redundant (\textit{i.e.}, it can be mapped to an equivalent gauge configuration inside this region, by means of consecutive infinitesimal gauge transformations). The proof of these features relies on two properties, namely: the hermiticity of the FP operator; and the existence of a minimizing functional defining the gauge fixing and the first Gribov region. In the Landau gauge, this functional is the norm $\frac{1}{2}\int_{x}  A_\mu^aA_\mu^a$ along the gauge orbit. Another gauge displaying these properties is the maximal Abelian gauge \cite{Capri:2008vk}, with the minimizing functional being $\frac{1}{2}\int_{x} A_\mu^iA_\mu^i$ with the index $i$ running only through the off-diagonal components of the gauge field. It turns out that, besides the fact that the LDW Faddeev-Popov operator is Hermitian, there is a minimizing functional defining the LDW gauge condition as well as the Gribov region, namely
\begin{equation}
\mathcal{F}[a]=\frac{1}{2}\int_{x} ~ \, a_\mu^aa_\mu^a\,.\label{min1}
\end{equation}
The first variation of \eqref{min1} along the gauge orbit leads to the gauge condition \eqref{ldwgaugecond}. The second variation and the requirement that the result is a minimum of \eqref{min1} leads to $\mathcal{M}^{ab}\ge 0$, which is the definition of the Gribov region in the LDW gauge. Thence, one has all the necessary elements to {check if} every configuration outside $\Omega$ is redundant. We leave that for future investigation.

 Once proven the fact that every gauge field configurations lying outside $\Omega$ can be continuously mapped to a configuration inside $\Omega$ (\textit{aka} gauge redundancy), the functional restriction to $\Omega$ should suffice to get rid of infinitesimal Gribov copies. Thus, in the search for gauge field configurations associated to positive eigenvalues of the FP operator, one may realize that such a chase is closely related to the analysis of the poles of the anti-ghost-ghost two-point function. From the ghosts action \eqref{ghaction} it is possible to see that the FP operator eigenvalues can be tracked down by the inverse of the anti-ghost-ghost propagator. Following \cite{Gribov:1977wm}, we will develop such investigation of the anti-ghost-ghost propagator at first order in perturbation theory.

As a consequence, one ends up with a consistency
condition imposed to the poles of the anti-ghost-ghost propagator, known as the
\textit{no-pole condition}. Such a condition will effectively impose the restriction to
$\Omega$, and with a new mass parameter {analogous to} the Gribov parameter in the Landau gauge that
must satisfy its own gap equation. 
Functionally, such a no-pole condition will be imposed by means of a modification to the functional measure
which, essentially, corresponds to the insertion of a Heaviside step function $\theta (x)$. Schematically,
\begin{align}
\label{gribovpathint}
    Z_{G} = 
    \mathcal{N}\int[da]\, \theta(\text{\it npc})\,
    \exp\left(-S[\bar{A},a] \right)
    \,,
\end{align}
with $S = S_{\text{YM}} + S_{gf} + S_{gh}$ and $npc$ stands for the no-pole condition to be imposed.

\section{The gap equation and the effective action}\label{nopolegapeq}
The first goal of this section is to investigate, up to one-loop order, the
no-pole condition {for} the anti-ghost-ghost propagator in the LDW gauge, by
following Gribov's prescription\footnote{We refer the reader to \cite{Sobreiro:2005ec} for a detailed derivation of the no-pole condition in the Landau gauge.} \cite{Gribov:1977wm}. In the sequence, the
Gribov restriction is implemented, so that one ends up with the
corresponding one-loop gap equation. Our second goal, which is the central point of
this paper, is to derive the effective action at leading order to verify its
background dependence, within Gribov's prescription.

\subsection{The no-pole condition and gap equation}
\label{subsectionghostformfactor}

According to \cite{Gribov:1977wm}, the no-pole condition is related to the trace
of the two-point Green's function of the ghost fields, while keeping the
fluctuation gauge field $a_{\mu}$ as {an external} field\footnote{Divergences appearing in the diagrams will be regularized by means of the dimensional regularization prescription. Therefore, arbitrary dimension will be denoted by $d$.}:
\begin{align}
\frac{\delta^{ab}}{V(N^{2}-1)}
\left\langle  \bar{c}^{a}(p) c^{b}(-p)   \right\rangle  \! [\bar{A},\,a]
~=~
{\cal G}[p^2,\,\bar{A},\,a]
\,.
\end{align}
In Section \ref{2pointgreensfunct} the fluctuation gauge field will be
quantized in order to actually compute the gauge and ghosts two-point
Green's function.

Namely, we will compute the following object, 
\begin{align}
\label{ahsgi}
{\cal G}[p^2,\,\bar{A},\,a]
~=~
\frac{1}{p^{2}}
\left[ 
1 + F(p^2,\bar{A},a)	
+ \sigma(p^2,\bar{A},a) 
\right]
\,.
\end{align}
The ghost form factor, in \eqref{ahsgi}, is split up into two factors, $F(p^2,\bar{A},a)$ and
$\sigma(p^2,\bar{A},a)$, denoting, respectively, the factor that does not
contribute to non-trivial poles ($p^2 \neq 0$) and the factor that actually
does. In other words, ${(1/p^2)F} \sim 1/p^4$ as we shall see,
and the positivity of the FP operator will be determined by $\sigma$. These
factors will be computed up to {leading} order, while keeping the fluctuation
field $a_{\mu}$ as an external field.

\begin{figure*}[t!]
\begin{center}
\includegraphics[width=1.0\textwidth,angle=0]{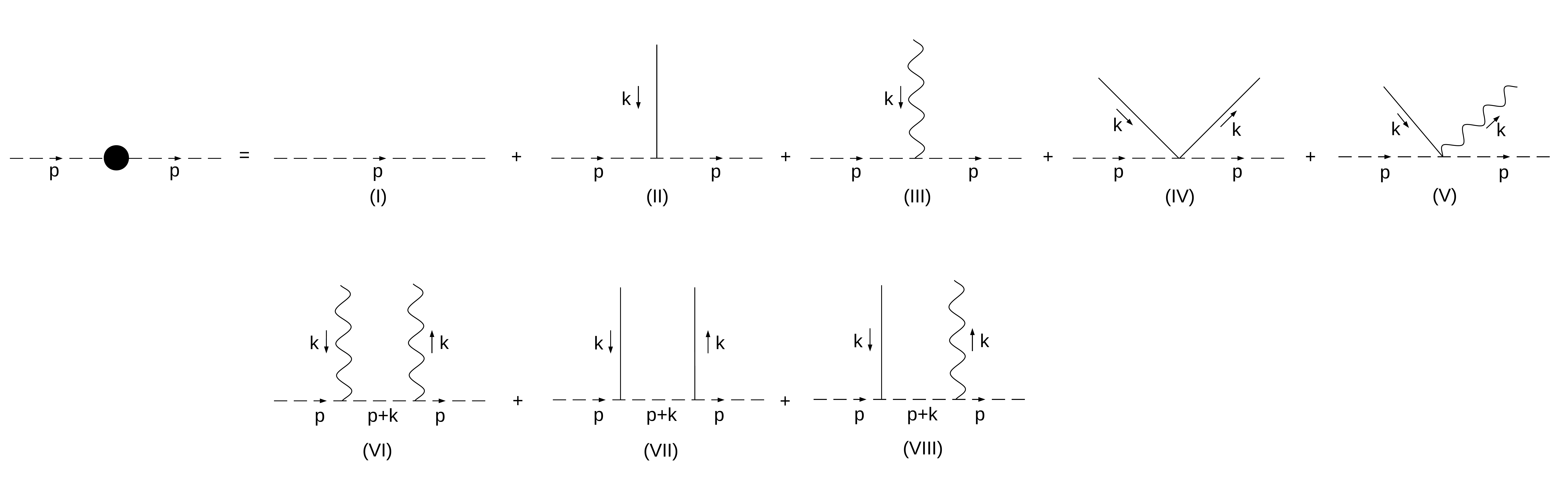}
\caption{
Anti-ghost-ghost two-point Green's function as a function of the external momentum $p^{2}$, the background {$\bar{A}_{\mu}$}, and of the fluctuation $a_{\mu}$ - which is taken as an external field for this calculation. Diagrams with linear dependence on the fluctuation field, $a_{\mu}$, are kept, although they do not contribute {to} the vacuum expectation value.}
\label{ghost_diag}
\end{center}
\end{figure*}

One can read off the one-loop contributing diagrams from the ghost action,
\begin{align}
\label{aeghilrg}
S_{gh} 
&=
\int_{x}\, 
\bigg\{
\bar{c}^{a} \p^{2} c^{a} 
+ 
\nonumber \\
&
+
gf^{ace} 
\big[ 
(\p_{\mu} \bar{A}^{e}_{\mu}) \bar{c}^{a} c^{c}
+
2\bar{A}^{e}_{\mu} (\p_{\mu}\bar{c}^{a}) c^{c}
+
a^{e}_{\mu} ( \p_{\mu}\bar{c}^{a} ) c^{c}
\big]
\nonumber \\
&
-
g^{2}f^{ace}f^{abd}
\big[
\bar{c}^{b}\bar{A}^{d}_{\mu}\bar{A}^{e}_{\mu} c^{c}
+
\bar{c}^{b} \bar{A}^{d}_{\mu} a^{e}_{\mu}c^{c}
\big]
\bigg\}
\,,
\end{align}
within perturbation theory. The one-loop two-point function of the ghost fields (with $a_{\mu}$ as an external field) is diagrammatically depicted in Figure \ref{ghost_diag} (\emph{cf}., Appendix \ref{thefeynmanrules} for the Feynman rules of the model.). Diagrams (II) and (III) do not contribute to the propagator for two reasons: by conservation of energy; and by taking the trace {in} the color space. 
Notice that the diagrams linear in $a_{\mu}$ are kept, although they do not contribute to any two-point Green function.



In order to derive the explicit expression of the diagrams of Figure 
\ref{ghost_diag} the background was considered to be transverse, $\p_{\mu} \bar{A}_{\mu} =0$ and independent of negative powers of the coupling constant. Furthermore,
the LDW gauge condition \eqref{ldwgaugecond} implies that
$\p_{\mu}a_{\mu}^{a} = g f^{abc}\bar{A}_{\mu}^{c}a_{\mu}^{b}$,
which means that the fluctuation $a_{\mu}$ can be considered transverse at zeroth order in the coupling constant $g$ 
(see the full gauge field two-point function in Appendix \ref{ghostdiagrams})\footnote{{The transversality of $a_{\mu}$ at $0^{\text{th}}$ order in $g$ relies on the assumption that $\bar{A}_{\mu}$ must be independent of $g^{-n}$ with $n>0$.}}.
Therefore, at arbitrary dimension we have
\begin{align}
\label{transfluct}
a_{\mu}^{a}(k) a_{\nu}^{b}(-k) =
\delta^{ab} 
\frac{1}{d-1} a_{\lambda}^{a}(k)  a_{\lambda}^{a}(-k) \; T_{\mu\nu}
\end{align}
and
\begin{align}
\bar{A}_{\mu}^{a}(k) a_{\nu}^{b}(-k) = \delta^{ab}
\frac{1}{d-1} \bar{A}_{\lambda}^{a}(k) a_{\lambda}^{a}(-k) 
\; T_{\mu\nu}
\,,
\end{align}
with $T_{\mu\nu}$ standing for the transverse projector, 
\begin{align}
T_{\mu\nu} = \delta_{\mu\nu} - \frac{k_{\mu}k_{\nu}}{k^{2}}
\,.
\label{top}
\end{align}
Thence, the two-point function ${\cal G}[p^2,\,\bar{A},\,a]$, given by equation \eqref{ahsgi}, can be rewritten as
\begin{align}
\label{rpoihgphg}
{\cal G}[p^2,\,\bar{A},\,a]
&~=~
\frac{1}{p^2}
\bigg[
1
+
F_{\textbf{IV}}(p^{2},\bar{A},a) 
+
\nonumber \\
&
+
F_{\textbf{V}}(p^{2},\bar{A},a) 
+
\sigma_{\textbf{VI}}(p^{2},\bar{A},a)
+
\nonumber \\
&
+
\sigma_{\textbf{VII}}(p^{2},\bar{A},a) 
+
\sigma_{\textbf{VIII}}(p^{2},\bar{A},a) 
\bigg]
\,,
\end{align}
with each term explicitly written {in} the Appendix \ref{thefeynmanrules}. {In particular}, the $\sigma$-terms will be treated collectively and denoted by $\sigma(p^{2},\bar{A},a)$, just as in {eq.}\eqref{ahsgi}:
\begin{align}
&
\sigma(p^{2},\bar{A},a) 
=
\nonumber \\
&=
\frac{g^2 N}{4V(d-1)(N^{2}-1)}
\frac{p_{\mu}p_{\nu}}{p^{2}}
\int_{k^{d}}
\Bigg[
\frac{ a^{d}_{\mu}(k) a^{d}_{\mu}(-k) }{(p+k)^{2}}    
\nonumber \\
&
+
2 \frac{ \bar{A}_{\mu}(k) a_{\mu}(-k)  }{(p+k)^{2}}
+
4 \frac{ \bar{A}^{d}_{\mu}(k)   \bar{A}^{d}_{\mu}(-k) }{(p+k)^{2}}
\Bigg] T_{\mu\nu}
\,.
\label{ghostformfactor1}
\end{align}
The $F$-terms will be collectively treated as 
\begin{align}
\label{e;ghgiwhe}
F(p^{2},\bar{A},a) 
&=
\frac{g^{2}N}{V(N^{2}-1)} 
\frac{1}{p^{2}}
\int_{k^{d}}
\bigg[
\bar{A}^{a}_{\mu}(k)\bar{A}^{a}_{\mu}(-k)
+
\nonumber \\
&
+
\int_{k^{4}}
\bar{A}^{a}_{\mu}(k)a^{a}_{\mu}(-k)
\bigg]\;.
\end{align}

From \eqref{ghostformfactor1} and \eqref{e;ghgiwhe} one is able to see that 
the two-point function \eqref{rpoihgphg} is a monotonically decreasing function
of $p^2$, so that its highest value is obtained at $p^2 =0$. Nonetheless, as can be
seen form \eqref{e;ghgiwhe} the diagrams of $F(p^2, A, a)$ \textit{only}
contribute to the pole of ${\cal G}[p^2,\,\bar{A},\,a]$ at $p^{2} =0$, and this
is why those diagrams will not be {taken into account} in the ghost form factor in
Gribov's prescription \footnote{Remember that the Gribov approach 
relies on the assumption that, in the thermodynamic limit, the only allowed pole for the anti-ghost-ghost propagator is at $p^{2}=0$, 
\cite{Gribov:1977wm,Vandersickel:2012tz}.}. Such strategy in dealing with the no-pole condition was first employed in the MAG, \cite{Capri:2005tj}. We also note that these diagrams are not well-behaved at the limit $p^2\rightarrow 0$, which will be taken bellow. Thence, the possibility of their disregard avoids external momentum dependence on the implementation of Gribov's restriction.

According to Gribov, \cite{Gribov:1977wm}, ${\cal G}[p^2,\,\bar{A},\,a]$ can be seen as a first-order approximation of 
\begin{align}
\label{traceghostprop0}
&{\cal G}[p^{2},\,\bar{A},\,a]
~\approx~
\frac{1}{p^{2}}
\frac{1}{\left[  1 - \sigma(p^{2},\bar{A},\,a)  \right]  }
+ 
\nonumber \\
&
+
\frac{1}{p^{2}}  F_{\textbf{IV}}(p^{2},\bar{A},a)
+
\frac{1}{p^{2}}  F_{\textbf{V}}(p^{2},\bar{A},a)
\,.
\end{align}
Thus, since the goal is to avoid poles other than $p^{2}=0$ of ${\cal G}[p^2,\,\bar{A},\,a]$, 
and $\sigma(p^{2},\bar{A},a)$ is a decreasing function of $p^{2}$, then it suffices to ensure that
\begin{equation}
\label{nopolecond0}
\sigma(0,\bar{A},\,a) < 1
\,,
\end{equation}
what has being called {as} \textit{Gribov's no-pole condition}.

As will become clear, in the thermodynamic limit where the microcanonical and the Boltzmann canonical ensembles become equivalent, \cite{Dudal:2009xh}, such no-pole condition boils down to the identity
\begin{equation}
\label{nopolecond1}
\sigma(0,\bar{A},\,a) = 1
\,.
\end{equation}
{The} factor $\sigma(0,\bar{A},a)$ computed at one-loop order reads
\begin{align}
\label{sig0}
    \sigma(0,\bar{A},a) 
    &= 
    \frac{g^{2}N}{Vd(N^{2}-2)}
    \int_{k^{d}}
    \Bigg[
    \frac{a_{\mu}^{a}(k)a_{\mu}^{a}(-k)}{k^{2}}
    \nonumber \\
    &
    + 2 \frac{\bar{A}_{\mu}^{a}(k)a_{\mu}^{a}(-k)}{k^{2}}
    + 4 \frac{\bar{A}_{\mu}^{a}(k)\bar{A}_{\mu}^{a}(-k)}{k^{2}}
    \Bigg]
    \;.
\end{align}

\subsubsection{The gap equation}

Before taking the thermodynamic limit, Gribov's no-pole condition, \eqref{nopolecond0}, can be functionally implemented by means of introducing the Heaviside step function, which leads us to Gribov's {partition} function,
\begin{align}
    \bar{Z}_{G} = \int[d\phi] \theta\left( 1 - \sigma(0,\bar{A},\,a) \right) \e^{-S[\bar{A},\phi]}
    \,,
    \label{Za}
\end{align}
with $S[\bar{A},\phi] = S_{\text{YM}} + S_{gf} + S_{gh}$, which is given by the equations \eqref{ym1}, \eqref{gf1} and \eqref{ghaction}, and 
$\phi = a_{\mu}^{a},\, \bar{c}^{a},\, c^{a}$. With the integral representation of the $\theta$-function, one can write the generating functional \eqref{Za} as,
\begin{align}
\bar{Z}_{G} [J] 
&= 
\int [d\phi] \int \frac{d\beta}{2\pi i \beta}
\,
\exp \bigg\{
-S[\bar{A},\phi] 
+
\nonumber \\
&
+
\beta(1-\sigma(0,\bar{A},a)) 
- 
\int_{k^{4}} J^{i}(k) \phi_{i}(k) 
\bigg\}
\,,
\label{Gribovpfunction}
\end{align}
where we introduced the sources $J^i$ associated to the field $\phi_i$ and the contraction $J^{i}\phi_{i}$ indicates a sum {over} the $i$-index that is counting for every source of each quantum field: $J_{\mu}^{a}a_{\mu}^{a}$; $\bar{J}_{c}^{a}c^{a}$; and $\bar{c}^{a}J_{\bar{c}}^{a}$. 


Considering only quadratic terms in the fields, the action $S[\bar{A},\phi]$ reads
\begin{align}
S[\bar{A},\phi]
&=
\int_{k^{4}} \frac12 
\bigg[
a_{\mu}^{a}(k) Q_{\mu\nu}^{ab} a_{\nu}^{b}(-k)
+
2 \bar{A}_{\mu}^{a}(k) Q_{\mu\nu}^{ab} a_{\nu}^{b}(-k)
\nonumber \\
&
+
\bar{A}_{\mu}^{a}(k) Q_{\mu\nu}^{ab} \bar{A}_{\nu}^{b}(-k)
+
\bar{c}^{a}(k) P^{ab} c^{b}(-k)
\bigg]
\nonumber \\
&
+
\frac{3g^{2}N \beta }{2V(N^{2}-1)}
\int_{k^{4}}
\frac{\bar{A}_{\mu}^{a}(k) \bar{A}_{\mu}^{a}(-k)}{k^{2}}
\,,\label{ZG2}
\end{align}
with $ P^{ab} = \delta^{ab} k^{2}$ and
\begin{align}
\label{Q}
Q_{\mu\nu}^{ab} 
&=
\left[
\left( k^{2} + \frac{g^{2}N \beta}{2V(N^{2}-1)} \frac{1}{k^{2}} \right) \delta_{\mu\nu}
\right.
\nonumber \\
&
\left.
-
\left( 1 - \frac{1}{\alpha} \right) \frac{k_{\mu}k_{\nu}}{k^{2}}
\right] \delta^{ab}
\,.
\end{align}
In order to write the action \eqref{ZG2} we considered that the background is transverse, \textit{i.e.} $k_{\mu}\bar{A}_{\mu}=0$, so that we could freely add the term $\frac{1}{2\alpha}\frac{k_{\mu}k_{\nu}}{k^{2}}\bar{A}_{\mu}$. Besides, we conveniently factored out the last term of \eqref{ZG2}.

By redefining the source
$
J_{\nu}^{b} = \tilde{J}_{\nu}^{b}(k)  - \bar{A}_{\mu}^{a}Q_{\mu\nu}^{ab}
\,,
$
one is able to integrate out the quantum fields in the partition function \eqref{Gribovpfunction} with the action \eqref{ZG2}. Namely, we have,
\begin{align}
    \bar{Z}_{G} [\tilde{J}] ~=~ \int \frac{d\beta}{2\pi i} \e^{-f(\beta)}
    \,.
\end{align}
In the thermodynamic limit the $\beta$-integral converges to the function $\e^{-f(\beta^{*})}$ with $\beta^*$ satisfying the saddle point equation, which is the the Gribov gap equation\footnote{\textit{Cf.,} \cite{Dudal:2009xh} for more details concerning {the} physical and mathematical aspects of the thermodynamic limit.},
\begin{align}
\label{hegihwe}
\frac{d f(\beta)}{d\beta} \Bigg\vert_{\beta = \beta^{*},\, J=0} = 0
\,.
\end{align}
The explicit expression of the gap equation reads,
\begin{align}
\label{gapequation3}
    \frac{(d-1)g^{2} N}{Vd}
    \int_{k^{d}}
    \frac{1}{k^{4} + \gamma^{* 4}}
    = 
    1
    -\bar{\sigma}[\bar{A}]
    \,,
\end{align}
with $\bar{\sigma}[\bar{A}]$ a constant (functional of $\bar{A}_{\mu}$) defined by
\begin{align}
    \bar{\sigma}[\bar{A}] = 
    \frac{4g^{2} N}{Vd(N^{2}-1)} \int_{k^{d}}
    \frac{\bar{A}_{\mu}^{a}(k)\bar{A}_{\mu}^{a}(-k)}{k^{2}}
\,.
\label{sigmabarra}
\end{align}
Besides, in the equation \eqref{gapequation3} we have defined $\gamma^{* 4} = \frac{2g^{2}N \beta^{*}}{d(N^{2}-1)}$ in order to simplify the notation, keeping in mind that here $\beta^{*}$ stands for the intensive Gribov parameter that solves equation \eqref{gapequation3}. It is then clear that the Gribov parameter is not free but depends of $g$. Moreover, at one-loop, there is a background-dependent contribution arising from eq.\eqref{sigmabarra}. For backgrounds with an intrinsic mass scale, this integral is not necessarily vanishing. Thus, the Gribov parameter, which is taken as a physical parameter in the standard RGZ setup, receives non-physical contributions from the background field already at one-loop order. This is related to the already reported issues reported in \cite{Dudal:2017jfw}.

Finally, the function $f(\beta^*)$ can be identified with the generating functional of connected Green's functions $\bar{W}[\bar{A},J]$,
\begin{align}
\label{gapeq0}
\bar{Z}_{G} [\tilde{J}] 
~=~
\int \frac{d\beta}{2\pi i}
\e^{-f(\beta)}
\approx
\e^{-f(\beta^{*})} = 
\e^{-\bar{W}[\bar{A},\tilde{J}]}\,,
\,
\end{align}
that explicitly reads,
\begin{align}
\label{liwerh}
\bar{W}[\bar{A},\tilde{J}] 
&=
\frac12 \int_{k^{4}}
\bigg[
\bar{A}_{\mu}^{a}(k) Q_{\mu\nu}^{ab} \bar{A}_{\nu}^{b}(-k)
+
\nonumber \\
&
+ \tilde{J}_{\mu}^{a}(k) {Q^{-1}}_{\mu\nu}^{ab} \tilde{J}_{\nu}^{b}(-k)
\bigg]
\nonumber \\
&
+ \frac12 \ln \det Q_{\mu\nu}^{ab}
+ \beta^{*} \bigg(\frac{3}{2}\bar{\sigma}[\bar{A}]
- 1 \bigg)
\nonumber \\
&
+ [gh. contrib.]
\,,
\end{align}
with $[gh. contrib.]$ standing for 
\begin{align}
\int_{k^{4}}  \tilde{J}_{\bar{c}}^{a}(k) {P^{-1}}^{ab} \tilde{J}_{c}^{b}(-k) - \ln\det P^{ab}
\,.
\label{ghcontrib}
\end{align}

It must be clear that the one-loop gap equation derived here, in the Yang-Mills
theory in the BFM, differs from the usual one (with the {trivial} background {$\bar{A}_\mu=0$}) by a constant that depends 
on the configuration of the background. 
Such {a} constant may be something quite intricate to be
computed, depending on the analytic expression of the background.
For the particular case where $\bar{A}_{\mu}^{a}$ is in a constant
configuration, such as in the Cartan sub-algebra of $SU(2)$, {which} is quite
useful in YM theory at finite temperature, 
\cite{Dudal:2017jfw,Reinosa:2014ooa,Reinosa:2014zta,Canfora:2019sen}, the
contribution to the gap equation is identically zero by dimensional regularization.

\subsection{The effective action}
\label{subsectioneffectiveaction}

To evaluate the effective action, it is convenient to rewrite
down the connected generating functional, \eqref{liwerh}, in terms of the (tree-level) connected
vacuum expectation value (v.e.v.) of the quantum fields $\phi_{c}$: 
\begin{align}
\label{conna}
{a_{c}}_{\mu}^{a}(k) &= \frac{\delta \bar{W}[J]}{\delta \bar{J}_{\mu}^{a}(k)} = - Q_{\mu\nu}^{-1 \; ab} \bar{J}_{\nu}^{b}(k) \,,
\\
\label{conncbar}
\bar{c}_{c}{}_{\mu}^{a}(k) &= \frac{\delta \bar{W}[J]}{\delta J_{\bar{c} \; \mu}^{a}(k)} = -\frac12 P^{-1 \; ab} J_{c}^{b}(k) \,,
\\
\label{connc}
{c_{c}}_{\mu}^{a}(k) &= \frac{\delta \bar{W}[J]}{\delta J_{c \; \mu}^{a}(k)} = - \frac12 J_{\bar{c}}^{b}(k) P^{-1 \; ba} \,,
\end{align}
where the subscript ``$c$'' stands for ``connected v.e.v.''.

The connected generating function can, then, be rewritten as
\begin{align}
\label{liwerhebf}
\bar{W}[J] 
&=
\frac12 \int_{k^{4}}
\Big[ \bar{A}_{\mu}^{a}(k) + {a_{c}}_{\mu}^{a}(k) \Big] Q_{\mu\nu}^{ab} \left[ \bar{A}_{\nu}^{b}(-k) + {a_{c}}_{\nu}^{b}(-k) \right] 
\nonumber \\
&
+ \frac12 \ln \det Q_{\mu\nu}^{ab}
+ \ln \beta^{*}
+ \beta^{*} \bigg(\frac{3}{2}\bar{\sigma}[\bar{A}] -1 \bigg)
\nonumber \\
&
+ [gh.contrib.]
+ \int_{k^{4}}
J^{i}(k) \phi_{c}{}_{i}(k)
\,,
\end{align}
where we used $J_{\nu}^{b} = \tilde{J}_{\nu}^{b}(k)  - \bar{A}_{\mu}^{a}Q_{\mu\nu}^{ab}$, and the equations 
\eqref{conna} -- \eqref{connc} in order to
derive the following relation
\begin{align}
- \tilde{J}_{\mu}^{a}(k) {Q^{-1}}_{\mu\nu}^{ab} \tilde{J}_{\nu}^{b}(-k) 
&=
2\bar{A}_{\mu}^{a} Q_{\mu\nu}^{ab} {a_{c}}_{\nu}^{b} 
+
{a_{c}}_{\mu}^{a} Q_{\mu\nu}^{ab} {a_{c}}_{\nu}^{b}
\nonumber \\
&
+
2 J_{\mu}^{a} {a_{c}}_{\mu}^{a} 
\,.
\end{align}
An equivalent manipulation was carried out on the ghost sector, $[gh.contrib.]$ in equation \eqref{ghcontrib}.

Finally, given the formal definition of the effective action,
\begin{align}
\bar{\Gamma}[\bar{A},a_{c}] = 
\bar{W}[J]
- \int_{k^{4}}
J^{i}(k)\phi_{c}{}_{i}(k)
\,,
\end{align}
one is able to derive {its explicit } expression {at one-loop order} in the BFM,
\begin{align}
&
\bar{\Gamma}[\bar{A},a_{c}]=
\nonumber \\
&=
\frac12 \int_{k^{4}}
\Big[ \bar{A}_{\mu}^{a}(k) + {a_{c}}_{\mu}^{a}(k) \Big] Q_{\mu\nu}^{ab} \left[ \bar{A}_{\nu}^{b}(-k) + {a_{c}}_{\nu}^{b}(-k) \right] 
\nonumber \\
&
+ \frac12 \ln \det Q_{\mu\nu}^{ab}
+ \beta^{*}\bigg(\frac{3}{2} \bar{\sigma}[\bar{A}] - 1\bigg)
\nonumber \\
&
+ [gh.contrib.]
\,.
\end{align}

Notice that, at {one-loop order} {for the vacuum energy}, we could explicitly verify that the effective action can be rewritten in terms of the ``total'' gauge field $A_{\mu} = \bar{A}_{\mu}+ a_{\mu}$ as follows,
\begin{align}
\bar{\Gamma}[\bar{A},a_{c}] 
&= 
\Gamma[\bar{A} + a_{c}] 
+ \frac{3\beta^{*}}{2}\bar{\sigma}[\bar{A}]
\nonumber \\
&=
\Gamma[A]
+ \frac{3\beta^{*}}{2}\bar{\sigma}[\bar{A}]
\,,
\label{effctiveaction}
\end{align}
where $\Gamma$ stands for the effective action obtained {in} the absence of the background. 

Therefore, it is immediate to verify that at {one-loop, the vacuum energy} of
Yang-Mills theory in the BFM, within Gribov's prescription, is \textit{not} background independent 
by a constant term proportional to the Gribov parameter $\beta^{*}$. Hence, it is clear that the background independence violation
is a direct consequence of Gribov's procedure, at least at {leading} order. Of course, that may not be the case at
higher loops (or may be, which would be in agreement with \cite{Dudal:2017jfw,Kroff:2018ncl}), and that is a matter of future investigation.

In the next section the propagator of the gauge and ghost fields {is} derived. In particular, the IR behavior of the ghost field propagator will be investigated at one-loop.

\section{The two-point Green's functions}
\label{2pointgreensfunct}

In this section we investigate the effects on the propagators of the gauge and
(anti-)ghost fields, due to the restriction of the path integral to the Gribov region, within the Yang-Mills BFM.

Since the zero modes of the FP operator (and therefore infinitesimal Gribov copies) seem to be generated at sufficiently large values of $g\, a_\mu$, it is expected that their removal will bring non-perturbative effects to the dynamics of Yang-Mills theories even if a perturbative expansion is performed.

The procedure developed in this section is the one commonly used in works devoted to the investigation of Gribov ambiguities (see \cite{Vandersickel:2012tz} for a detailed review and references therein). In summary, one should compute the inverse of $Q^{ab}_{\mu\nu}$ in order to derive the gauge field two-point Green's function, and compute $\left\langle \sigma(p^{2},\bar{A},a)  \right\rangle$ and $\left\langle F(p^{2},\bar{A},a)  \right\rangle$ with equations \eqref{ghostformfactor1} and \eqref{e;ghgiwhe}, respectively, in order to derive the ghost two-point Green's function. At this point, one should keep in mind that $\bar{A}_{\mu}$ is a {classical} field, so that $\left\langle \bar{A}_{\mu}  \right\rangle = \bar{A}_{\mu}$, and that tadpole-like diagrams (those with just one insertion of $a_{\mu}$) vanish.

The investigation of the IR physics of the gauge and (anti-)ghost
propagators
has been the subject of numerous works, {ranging} from {applications of} Dyson-Schwinger {equation},
\cite{Cornwall:1981zr,Alkofer:2000wg,Fischer:2006ub,Aguilar:2008xm,Alkofer:2008jy}, functional renormalization group \cite{Cyrol:2016tym} to
{lattice} quantum field theory
\cite{Oliveira:2007dy,Cucchieri:2007rg,Bogolubsky:2009dc,Cucchieri:2010xr,Cucchieri:2016jwg}.

\subsection{The gauge {field} propagator}

The connected two-point Green's function, at tree-level, of the ({fluctuating}) gauge field can
be {easily} read off {from} equation \eqref{liwerh}. In summary, one should compute
the inverse of the $Q_{\mu\nu}$ operator, which reads, 
\begin{eqnarray}
    \left( Q^{-1} \right)^{ab}_{\mu\nu} &=&
    \delta^{ab} \left(
    \frac{k^{2}}{k^{4} + \gamma^{\ast}{}^{4}} \delta_{\mu\nu}
    -
    \right.
    \nonumber\\ 
    &-&\left.\frac{k^{4}(1-\alpha)}{(k^{4} +
    \gamma^{\ast}{}^{4})(\alpha(k^{4}+\gamma^{\ast}{}^{4})-k^{2}\alpha + k^{2})}
\frac{k_{\mu}k_{\nu}}{k^{2}}
\right)
\,.\nonumber\\
\end{eqnarray}
Taking the LDW gauge limit, $\alpha \to 0$, we have 
\begin{align}
    \left( Q^{-1} \right)^{ab}_{\mu\nu} =
    \delta^{ab}
    \frac{k^{2}}{k^{4} + \gamma^{\ast}{}^{4}}
    \left(\delta_{\mu\nu} - \frac{k_{\mu}k_{\nu}}{k^{2}} \right)
\,,
\label{gaugeprop}
\end{align}
where terms proportional to $g^{2}$ have been neglected\footnote{A detailed discussion concerning the non-transversality feature of the gauge field propagator can be found in Appendix \ref{ghostdiagrams} and briefly above equation \eqref{transfluct}.}. 
In the above equation, \eqref{gaugeprop}, 
remind that $\gamma^{* 4} = \frac{2g^{2}N \beta^{*}}{d(N^{2}-1)}$, with $\beta^{*}$ the Gribov parameter that solves the gap equation, \eqref{gapequation3}.

Therefore, notice that the gauge field propagator, given by equation
\eqref{gaugeprop}, has the same tree-level expression as the one obtained in
the absence of the background,
\cite{Vandersickel:2012tz}. 
%
%
However, by solving the gap equation, \eqref{gapequation3}, one can verify that the 
Gribov parameter depends on the background configuration,
%
%
\begin{align}
\label{gribovbehavior}
    \beta^{*} = \frac{3}{g^{2}}\bar{\mu}^{2}\e^{\frac{1}{3}-
    \frac{32\pi^{2}}{3g^{2}}(1-\bar{\sigma}[\bar{A}])}
    \,,
\end{align}
which brings a possible background dependence to the propagator of the gauge field. Notice that the above expression for the Gribov parameter, 
\eqref{gribovbehavior}, differs from the usual result $\sim \e^{\frac{1}{3}-\frac{32\pi^{2}}{3g^{2}}}$ by the presence of the functional $\bar{\sigma}[\bar{A}]$.

From \eqref{gribovbehavior} one can foresee two particular scenarios that deserve special attention:

\begin{itemize}

\item[i.] $\bar{\sigma}[\bar{A}] = 0$

In this case the behavior of the {gauge field} propagator is equivalent to the usual
behavior obtained by Gribov in \cite{Gribov:1977wm}. Namely,
\begin{align}
\label{gribovbehaviorsigma0}
    \beta^{*} = \frac{3}{g^{2}}\bar{\mu}^{2}\e^{\frac{1}{3}-
    \frac{32\pi^{2}}{3g^{2}}}
    \,.
\end{align}
This specific scenario can be obtained for constant configurations for the background (making sure that $\bar{A}_{\mu}$ does not depend on non-negative powers of $g$). At the end of this section a few words are dedicated to the constant background configuration, where the vanishing of $\bar{\sigma}[\bar{A}]$ is investigated.

A particularly interesting constant configuration is the one where the background is set up in the $SU(2)$ Cartan sub-algebra, $\bar{A}^{a}_{\mu} = \bar{A}\delta^{3a}\delta_{0\mu}$ with $\bar{A}$ just a constant. Such a configuration is useful in the probe of the (de)confinement phase transition at finite temperature, since it can mimic the Polyakov loop, \cite{Reinosa:2014ooa,Reinosa:2014zta,Canfora:2015yia}.

\item[ii.]$\bar{\sigma}[\bar{A}] = 1$

In this case the Gribov parameter behaves like
\begin{align}
    \label{asghihg1}
    \beta^{*} = \frac{3}{g^{2}}\bar{\mu}^{2}\e^{\frac{1}{3}}
    \,,
\end{align}
{making} clear that the {behavior} of the Gribov parameter depends on the configuration of the background. Notice that in cases where $\beta^{*}$ behaves like \eqref{asghihg1} the Gribov parameter becomes {larger} in the deep UV regime (at least in the one-loop approximation). That is drastically different from the standard scenario given by \eqref{gribovbehaviorsigma0} but it is also a case which requires a lot of fine tuning in the background sector to exactly cancel the constant part of the argument of the exponential in \eqref{gribovbehavior}.

\end{itemize}

\subsection{Ghost propagator}

Let us now investigate the IR behavior of the anti-ghost-ghost two-point function at one-loo order\footnote{Notice that the tree-level ghost propagator is not changed by the Gribov prescription, so that one must go to the next order to see any effect.}. For that, consider the v.e.v. of ${\cal G}[p^{2},\bar{A},a]$, in equation \eqref{rpoihgphg}, where the Gribov no-pole condition is satisfied in order to validate the (first order) approximation,
\begin{align}
	\label{wigwg3123}
	\left\langle \bar{c}^{a}(p) c^{a}(-p) \right\rangle
	&
	~\approx~
	\frac{1}{p^{2}}
	\frac{1}{\left[  1 - \left\langle \sigma(p^{2},\bar{A},\,a) \right\rangle  \right]  }
	+ 
	\nonumber \\
	&
	\phantom{~\approx~}
	+\frac{1}{p^{2}} \left\langle F_{\textbf{IV}}(p^{2},\bar{A},a) \right\rangle 
	\,,
\end{align}
with $\left\langle  \sigma(p^{2},\bar{A}, a) \right\rangle$ given by
\begin{align}
	&
	\left\langle  \sigma(p^{2},\bar{A}, a) \right\rangle 
	= 
	\nonumber \\
	&=
	\frac{g^2 N}{V(d-1)(N^2-1)}
    \frac{p_{\mu}p_{\nu}}{p^{2}}
	\int_{k^{d}}
	\frac{k^{2}}{(p+k)^2} 
	\frac{T_{\mu\nu}(k)}{k^4 + \gamma^4}
	\nonumber \\
	&
	+ \frac{d}{d-1}  \frac{p_{\mu}p_{\nu}}{p^{2}} \bar{\sigma}_{\mu\nu}[\bar{A}]
	\,,
	\label{vevgff0}
\end{align} 
where we have defined
\begin{align}
    \bar{\sigma}_{\mu\nu}[\bar{A}] = 
    \frac{4g^{2} N}{Vd(N^{2}-1)} \int_{k^{d}}
    \frac{\bar{A}_{\alpha}^{a}(k)\bar{A}_{\alpha}^{a}(-k)}{k^{2}}
    T_{\mu\nu}
\,,
\end{align}
and
\begin{align}
\label{e;ghgiwhe2}
&
\left\langle F_{\textbf{IV}}(p^{2},\bar{A},a) \right\rangle 
=
\nonumber \\
&=
\frac{g^{2}N}{V(N^{2}-1)} 
\frac{1}{p^{2}}
\int_{k^{d}}
\bar{A}^{a}_{\mu}(k)\bar{A}^{a}_{\mu}(-k)
\,.
\end{align}

After some manipulations we get
\begin{widetext}
\begin{align}
	\label{wigwg312}
	\left\langle \bar{c}^{a}(p) c^{a}(-p) \right\rangle
	&~\approx~
	\frac{1}{p^{4}}
	\left(
	\frac{d(d-1)(d+2)}{ [d(d-3)+2](I^{d}_{\gamma}+I^{d}_{\bar{A}})  }
	+ 
 	\frac{g^{2}N}{V(N^{2}-1)} 
	\int_{k^{d}}
	\bar{A}^{a}_{\mu}(k)\bar{A}^{a}_{\mu}(-k)
	\right)
	\,,
\end{align}
\end{widetext}
with,
    \begin{align}
    \label{igamma}
    I^{d}_{\gamma} = 
    \frac{g^2 N}{V(N^2-1)}
	\int_{k^{d}}
	\frac{1}{k^{2}(k^4 + \gamma^4)}
	\,,
    \end{align}
    and
    \begin{align}
    \label{iabarra}
    I^{d}_{\bar{A}} = 
    \frac{4g^2 N}{V(N^2-1)}
	\int_{k^{d}}
	\frac{\bar{A}_\mu^a(k) \bar{A}_\mu^a(-k)}{k^{4}}\,.
    \end{align}
Notice that the gauge field propagator \eqref{gaugeprop} was used to write down the equation \eqref{vevgff0}, as well as the gap equation \eqref{gapequation3}
was employed to derive \eqref{wigwg312}. At the end, with the factors $I^{d}_{\gamma}$ and $I^{d}_{\bar{A}}$ given,
respectively, by equations \eqref{igamma} and \eqref{iabarra}, one can verify
that the IR behavior of the ghost two-point function is enhanced 
($\sim p^{-4}$), regardless {of} the  background
configuration.

\subsection{The constant background}

As was shown in equation (\ref{effctiveaction}), the vacuum energy is background-dependent, by a constant term proportional to the Gribov parameter:
\begin{eqnarray}
  \mathcal{I} = 
  \frac{3\beta^{*}}{2}\bar{\sigma}[\bar{A}]
  \,,
  \label{breaking}
\end{eqnarray}
with $\bar{\sigma}[\bar{A}]$ given by equation \eqref{sigmabarra}. Let us investigate the constant background configuration.

When the background configuration is constant the function $\bar{\sigma}[\bar{A}]$ of equation \eqref{breaking} reduces to 
\begin{align}
    \bar{\sigma}[\bar{A}] = 
    \frac{4g^{2} N \bar{A}_{\mu}^{a}\bar{A}_{\mu}^{a} }{Vd(N^{2}-1)} 
    \int_{k^{d}}
    \frac{1}{k^{2}}
\,,
\label{sigmabarracte}
\end{align}
Since there is no mass scale involved in the integrand, the result must be purely infinite. That is, once the regularization scheme is chosen, the finite part of the result can be canceled. Particularly, this sort of divergent integral is zero by means of the dimensional regularization scheme.
\section{Conclusions}

In this paper we investigate the background dependence of Yang-Mills theories
at one-loop order within the Landau-DeWitt gauge by taking into account the existence of gauge copies. As a consequence of the
FP procedure to impose the LDW gauge condition, the existence of
zero modes of the FP operator was dealt with by following the original
perturbative approach of V.~N.~Gribov, \cite{Gribov:1977wm}, within the
the BFM. The background field, $\bar{A}_{\mu}$, was
considered to be transverse and independent of $g^{-n}$ (with $n > 0$), and the fluctuation field, $a_{\mu}$, could be
treated as transverse at $0^{\text{th}}$ order in the coupling constant expansion, as a consequence of the LDW condition combined with $\bar{A}_{\mu}$ depending on non-negative powers of $g$. At higher order of
loop corrections (or for $\bar{A}_{\mu} \propto g^{-1}$), longitudinal contributions of $a_{\mu}$ 
should be taken into account. After investigating which diagrams contribute to the
poles of the anti-ghost-ghost two-point Green's function, which is computed as a function of
$\bar{A}_{\mu}$ and $a_{\mu}$, the one-loop no-pole condition was determined,
as can be seen in equation \eqref{nopolecond0} with the ghost form factor given
in equation \eqref{ghostformfactor1}. As a consequence of imposing the no-pole
condition {to} the functional {measure} we could verify that the vacuum energy is
\textit{not} background independent anymore, in agreement with the works
\cite{Canfora:2015yia,Canfora:2016ngn,Dudal:2017jfw,Kroff:2018ncl}. At leading
order we could verify that the background independence of the effective action
is lost due to a constant factor $\bar{\sigma}[\bar{A}]$ that depends on the
configuration of $\bar{A}_{\mu}$. Namely, this functional reads, 
\begin{align}
    \label{barsigma1}
    \bar{\sigma}[\bar{A}] = 
    \frac{4g^{2} N}{Vd(N^{2}-1)} 
    \int_{k^{d}}
    \frac{\bar{A}_{\mu}^{a}(k)\bar{A}_{\mu}^{a}(-k)}{k^{2}}
\,.
\end{align}

An immediate side effect of the background dependence (at leading
order, at least) is that the behavior of the Gribov parameter strongly
depends on the configuration of the background. By solving the gap equation
\eqref{gapequation3} we could identify two important scenarios: 
\begin{itemize}

\item[i.]
if $\bar{\sigma}[\bar{A}] = 0$, where the Gribov parameter behaves just as in the
standard YM theory (\textit{i.e.} in the absence of the background).  This is
the case of constant configurations of the background, as \textit{e.g.} in the
Cartan subalgebra of $SU(2)$ settings, used by the authors of
\cite{Canfora:2015yia,Canfora:2016ngn,Dudal:2017jfw,Kroff:2018ncl}; 

\item[ii.]
if 
$\bar{\sigma}[\bar{A}] = 1$, where the Gribov parameter behaves drastically
different from the standard YM case. In this second scenario, $\beta$ becomes more
relevant in the UV regime, so that the usual perturbative features of the
standard YM theory cannot be recovered anymore.
\end{itemize}

Furthermore, We could also verify that whenever $\bar{A}_{\mu}^{a}(k)\bar{A}_{\mu}^{a}(-k)$
is a function of $k^{2}$ {without} a mass parameter, then the constant
$\bar{\sigma}[\bar{A}]$ is zero by dimensional regularization. The trivial case
is the constant configuration of the background, and in particular when the background is
in the Cartan subalgebra of $SU(2)$. Thus, at one-loop the constant background
configuration is a very special case where the vacuum energy
does not receive any contributions from the background. 

This background independence for the
special case of constant configurations of the background seems to be in
\textit{disagreement} with the results of
\cite{Canfora:2015yia,Canfora:2016ngn,Dudal:2017jfw,Kroff:2018ncl}. It must be
clear that here we are imposing a different (perturbative) no-pole condition.
In contrast to
\cite{Canfora:2015yia,Canfora:2016ngn,Dudal:2017jfw,Kroff:2018ncl}, we are
perturbatively building up the horizon function, which is given by means of the
ghost form factor in equation \eqref{sig0}, rather than proposing an \textit{ad
hoc} expression (either perturbative \text{\`a la} Gribov
\cite{Canfora:2015yia,Canfora:2016ngn} or at all orders \text{\`a la} Zwanziger
\cite{Dudal:2017jfw,Kroff:2018ncl}) with the compromise of being able to
recover the usual results in the absence of the background. However, we have no particular reason to expect that at higher orders, a non-trivial background-dependent contribution to the vacuum energy will not arise and therefore, we do expect that background independence will be spoiled by the restriction to the Gribov region even for a constant background.


It must be clear that our approach strongly relies on the assumption that the background field must be proportional to some non-negative power $g$. We pointed out that even the constant background configuration must satisfy such a condition.
The Henyey configuration, \cite{Henyey:1978qd}, is another possibility for the background, which satisfy the hypothesis that $\bar{A}_{\mu}$ is independent of $g^{-n}$, $n>0$. Such a configuration is of particular interest in the study of zero modes of the FP operator in the Landau gauge, \cite{Sobreiro:2005ec,Capri:2012ev}. The investigation of Henyey's configuration within the BFM is the subject of future work. In principle, one must be careful when considering instanton configurations, since it may violate the assumption that $\bar{A}_{\mu}$ is independent of $g^{-n}$. A suitable YM-BFM approach for instantons is currently being developed.

Our results strongly relies on the assumption that all gauge field
configuration that lies outside the Gribov region $\Omega$ (given {by} equation
\eqref{omegadefinition}) can be mapped to a gauge field configuration within
$\Omega$ by successive infinitesimal gauge transformations. Further investigation to underpin
such {a} statement is a matter of future work, along side with a recursive
construction of an all order Gribov horizon, \textit{\`a la}
\cite{Capri:2012wx}, as well as the investigation of renormalization aspects
of this model.

\section*{Acknowledgments}
ADP acknowledges CNPq under the grant PQ-2 (309781/2019-1) and FAPERJ under the
``Jovem Cientista do Nosso Estado" program (E-26/202.800/2019). The Coordena\c
c\~ao de Aperfei\c coamento de Pessoal de N\'ivel Superior (CAPES) is also
acknowledged for financial support. I.F.J. acknowledges CAPES for the financial
support under the project grant $88887.357904/2019-00$.

\appendix

\section{Hermiticity of the FP operator}\label{Ap1}

The Gribov method for eliminating copious configurations relies on the restriction of the functional integration to the so called first Gribov region, which is defined as the set of all gauge field configurations for which the FP operator $\mathcal{M}$ is strictly positive. More precisely, if

\begin{eqnarray}
  \mathcal{M}^{ab}\Tilde{A}_{\mu}^{b} = \omega \Tilde{A}_{\mu}^a,
\end{eqnarray}
for some positive and real eigenvalue $\omega,$ then we say that the
configuration $\tilde{A}_{\mu}^{a}$ belongs to the first Gribov region. It is,
therefore, crucial that $\mathcal{M}$ be a hermitian operator, {within Gribov and Zwanziger's approach
\cite{Gribov:1977wm,Zwanziger:1989mf,Vandersickel:2012tz}}.

It shall be proven in what follows that, indeed, $\mathcal{M}$ is Hermitian in
the LDW gauge. We first note that the operator in (\ref{fpop}) can be written as
\begin{eqnarray}
  \mathcal{M}^{ab} = -\bar{D}_{\mu}^{ac}\bar{D}_{\mu}^{cb}+gf^{cbd}\bar{D}_{\mu}^{ac}\left(a_{\mu}^{d}\cdot\right).
  \label{fpop2terms}
\end{eqnarray}
For this operator to be hermitian we must show that

\begin{eqnarray}
  \int_{x} \ \psi^{a}\mathcal{M}^{ab}\phi^{b}=
  \int_{x} \ \phi^{a}\mathcal{M}^{ab}\psi^{b}
\end{eqnarray}
for general Lie-algebra valued configurations $\psi$ and $\phi.$ One may easily verify that the first term of (\ref{fpop2terms}) is
Hermitian,
since it
is the product of two anti-Hermitian operators. More precisely, {such
non-Hermitian operator satisfy the following identity,}
\begin{eqnarray}
  \int_{x} \ \psi^a \bar{D}_{\mu}^{ab} \phi^b =
  -\int_{x} \ \phi^a \bar{D}_{\mu}^{ab} \psi^b.
  \label{DbarAntiHerm}
\end{eqnarray}

As for the second term in (\ref{fpop2terms}) one has
\begin{align}
  &
  \int_{x} \ \psi^{a}\bar{D}_{\mu}^{ac}\left(gf^{cbd}a_{\mu}^{d}\phi^{b}\right)
  =
  \nonumber \\
  &=
  -\int_{x} \ \bar{D}_{\mu}^{ca}\psi^{a}\left(gf^{cbd}a_{\mu}^{d}\phi^{b}\right),
\label{wgh2g12}
\end{align}
where use has been made of equation (\ref{DbarAntiHerm}). Expanding
$\bar{D}_{\mu}^{ac}$ in {\eqref{wgh2g12} lead
us to}
\begin{align}
\label{lgi4hgl2n1}
    &
    \int_{x} \
    \psi^{a}\bar{D}_{\mu}^{ac}\left(gf^{cbd}a_{\mu}^{d}\phi^{b}\right) =
    \nonumber \\
    &=
    \int_{x} \
    \phi^{b}
    \bigg[\partial_{\mu}\left(gf^{bcd}\psi^{c}a_{\mu}^{d}\right)
    +gf^{cbd}\psi^{c}\partial_{\mu}a_{\mu}^{d}
    \nonumber \\
    &
    \phantom{\int_{x} \ \phi^{b}\bigg[\partial_{\mu}}
    +g^{2}f^{cae}f^{cbd}a_{\mu}^{d}\psi^{a}\bar{A}_{\mu}^{e}
    \bigg]
\,,
\end{align}
where integration by parts was used. Now we use the definition of
$\bar{D}_{\mu}^{ab}$ and the LDW gauge condition to rewrite the {left-hand-side of equation \eqref{lgi4hgl2n1}}
%
\begin{align}
    &
    \int_{x} \ \phi^{a}\bar{D}_{\mu}^{ac}\left(gf^{cbd}\psi^{b}a_{\mu}^{d}\right)
    +\int_{x} \ \phi^{b}
    \bigg(g^{2}f^{bce}f^{cad}\psi^{a}a_{\mu}^{d}\bar{A}_{\mu}^{e}
    \nonumber \\
    &
    +g^{2}f^{abc}f^{cde}\psi^{a}a_{\mu}^{d}\bar{A}_{\mu}^{e}
    +g^{2}f^{cae}f^{cbd}\psi^{a}a_{\mu}^{d}\bar{A}_{\mu}^{e}
    \bigg)
\,,
\end{align}
where we have renamed dummy indices. Notice that the second integral vanishes by virtue of the Jacobi identity. We are, thus, left with
\begin{align}
    &
    \int_{x} \ \psi^{a}\bar{D}_{\mu}^{ac}\left(gf^{cbd}a_{\mu}^{d}\phi^{b}\right) =
    \nonumber \\
    &=
    \int_{x} \ \phi^{a}\bar{D}_{\mu}^{ab}\left(gf^{bcd}\psi^{c}a_{\mu}^{d}\right)
 \,,
\end{align}
just as we {meant} to show. This concludes the proof of the
Hermiticity of $\mathcal{M}$ given by  (\ref{fpop}). Hence, it makes sense
defining the region where $\mathcal{M} > 0.$



\section{The Feynman rules and diagrams}
\label{thefeynmanrules}

\subsection{The Feynman rules}

Here we present the Feynman rules, which can easily be read off the action \eqref{aeghilrg}, and that are useful to compute the diagrams that contribute to the one-loop ghost two-point function (and are depicted in the Figure \ref{ghost_diag}). Namely,
\begin{widetext}

\begin{eqnarray}
  \begin{tikzpicture}[baseline=(b.base),horizontal=a to b]
    \begin{feynman}
      \vertex (a) at (-1.2,0);
      \vertex (b) at (-1,0);
      \vertex (c) at (1,0);
      \vertex (d) at (1.2,0);
      \diagram{
        a [particle=\(\bar{c}^{a}\)] -- [scalar] b 
        -- [charged scalar, edge label'=\(p\)] c 
        -- [scalar] d [particle=\(c^{b}\)] ,
      };
    \end{feynman}
  \end{tikzpicture}
  &
  \quad = \quad 
  &
  \frac{1}{p^{2}}\delta^{ab}
\\
  \begin{tikzpicture}[baseline=(b.base),horizontal=a to b]
    \begin{feynman}
      \vertex (a) {\(\bar{c}^{a}\)};
      \vertex [right=of a] (b);
      \vertex [right=of b] (c) {\( c^{b} \)};
      \vertex [above=of b] (d) {\( \bar{A}_{\mu}^{d} \)};
      \diagram{
        (a) -- [charged scalar, edge label'=\(p\)] (b) 
        -- [charged scalar, edge label'=\(p\)] (c),
        (d) -- [ momentum=\(k\)] (b),
      };
    \end{feynman}
  \end{tikzpicture}
  &
  \quad = \quad 
  &
  igf^{ace}(2p + k)_{\mu} \delta^{4}(k)
  \end{eqnarray}
  \begin{eqnarray}
  \begin{tikzpicture}[baseline=(b.base),horizontal=a to b]
    \begin{feynman}
      \vertex (a) {\(\bar{c}^{a}\)};
      \vertex [right=of a] (b);
      \vertex [right=of b] (c) {\( c^{b} \)};
      \vertex [above=of b] (d) {\( \bar{A}_{\mu}^{d} \)};
      \diagram{
        (a) -- [charged scalar, edge label'=\(p\)] (b) 
        -- [charged scalar, edge label'=\(p\)] (c),
        (d) -- [photon, momentum=\(k\)] (b),
      };
    \end{feynman}
  \end{tikzpicture}
  &
  \quad = \quad 
  &
  i gf^{ace}p_{\mu} \delta^{4}(k)
  \end{eqnarray}
  \begin{eqnarray}
     \begin{tikzpicture}[
  arrowlabel/.style={
    /tikzfeynman/momentum/.cd, 
    arrow shorten=#1,arrow distance=1.5mm},
    arrowlabel/.default=0.4,
  baseline=(b.base),horizontal=a to b]
    \begin{feynman}
      \vertex (a) {\(\bar{c}^{a}\)};
      \vertex [right=of a] (b);
      \vertex [right=of b] (c) {\( c^{b} \)};
      \vertex [above left=of b] (d) {\( \bar{A}_{\mu}^{e} \)};
      \vertex [above right=of b] (e) {\( \bar{A}_{\nu}^{d} \)};
      \diagram{
        (a) -- [charged scalar, edge label'=\(p\)] (b) 
        -- [charged scalar, edge label'=\(p\)] (c),
        (d) -- [momentum={[arrowlabel]$k$}] (b),
        (e) -- [momentum'={[arrowlabel]$k$}] (b),
      };
    \end{feynman}
  \end{tikzpicture}
  &
  \quad = \quad 
  &
  - \frac12 g^{2}  \big( f^{abd}f^{ace} 
  \nonumber \\
  &&
  \phantom{- \frac12}
  +f^{abe}f^{acd}  \big) \delta_{\mu\nu}
  \end{eqnarray}
  \begin{eqnarray}
     \begin{tikzpicture}[
  arrowlabel/.style={
    /tikzfeynman/momentum/.cd, 
    arrow shorten=#1,arrow distance=1.5mm},
    arrowlabel/.default=0.4,
  baseline=(b.base),horizontal=a to b]
    \begin{feynman}
      \vertex (a) {\(\bar{c}^{a}\)};
      \vertex [right=of a] (b);
      \vertex [right=of b] (c) {\( c^{b} \)};
      \vertex [above left=of b] (d) {\( \bar{A}_{\mu}^{e} \)};
      \vertex [above right=of b] (e) {\( a_{\nu}^{d} \)};
      \diagram{
        (a) -- [charged scalar, edge label'=\(p\)] (b) 
        -- [charged scalar, edge label'=\(p\)] (c),
        (d) -- [momentum={[arrowlabel]$k$}] (b),
        (e) -- [photon, momentum'={[arrowlabel]$k$}] (b),
      };
    \end{feynman}
  \end{tikzpicture}
  &
  \quad = \quad 
  &
  - g^{2} f^{abd}f^{ace} \delta_{\mu\nu}
\end{eqnarray}

\end{widetext}

\subsection{The one-loop contributing diagrams}
\label{ghostdiagrams}

Here one may find the explicit expression of each diagram that contributes to the two-point Green's function of the ghost fields at one-loop, including those diagrams that are linear in $a_{\mu}$.

In order to derive each expression we made use of three important assumptions:
\begin{itemize}
    \item[1.] The background is supposed to be transverse,
    \[
    \p_{\mu}\bar{A}_{\mu} =0;
    \]

    \item[2.] The background field must \textit{not} depend on non-negative powers of $g$. That is, if
    \[
    \bar{A}_{\mu} \propto g^{n} \,
    \]
    then $n \geq 0$.

    \item[3.] The LDW gauge implies that $\p_{\mu}a_{\mu}^{a} = g f^{abc}\bar{A}_{\mu}^{c}a_{\mu}^{b}$ (with $\bar{A}_{\mu}$ satisfying the assumption 2 stated above), so that two-point composite operator $a_{\mu}^{a}(k)a_{\nu}^{b}(-k)$ can be decomposed as 
\begin{widetext}
\begin{align}
\label{transfluct1}
a_{\mu}^{a}(k) a_{\nu}^{b}(-k) 
&= 
\delta^{ab} 
\left[
\frac{1}{d-1} a_{\lambda}^{a}(k)  a_{\lambda}^{a}(-k) \; T_{\mu\nu}
+
\right.
\nonumber \\
&+
\left.
\frac{g^{2}}{k^{2}}f^{abc}f^{ade}\bar{A}_{\lambda}^{c}(k) \bar{A}_{\rho}^{e}(-k) a_{\lambda}^{b}(k) a_{\rho}^{d}(-k) \; \left( L_{\mu\nu} - \frac{1}{d-1}T_{\mu\nu} \right)
\right]
\,,
\end{align}
\end{widetext}
with $T_{\mu\nu}$ and $L_{\mu\nu}$ standing for the transversal and longitudinal projectors, 
\begin{align}
T_{\mu\nu} = \delta_{\mu\nu} - \frac{k_{\mu}k_{\nu}}{k^{2}}
\quad \text{and} \quad
L_{\mu\nu} = \frac{k_{\mu}k_{\nu}}{k^{2}}
\,.
\end{align}
Therefore, as mentioned in Section \ref{nopolegapeq}, the longitudinal contribution of $a_{\mu}^{a}(k)a_{\nu}^{b}(-k)$ does only contribute to the next order in perturbation theory. Thus, at first order one is allowed to take into account only the transverse contribution. Namely, 
    \begin{align*}
    a_{\mu}^{a}(k) a_{\nu}^{b}(-k) 
    = 
     \frac{\delta^{ab}}{d-1} a_{\lambda}^{a}(k)  a_{\lambda}^{a}(-k) \; 
     \left( \delta_{\mu\nu} - \frac{k_{\mu}k_{\nu}}{k^2}  \right) \,.
    \end{align*}

\end{itemize}

The diagrams are related bellow,

\begin{eqnarray}
\text{(IV)} 
\quad &=& \quad
  \begin{tikzpicture}[
  arrowlabel/.style={
    /tikzfeynman/momentum/.cd, 
    arrow shorten=#1,arrow distance=1.5mm},
    arrowlabel/.default=0.4,
  baseline=(b.base),horizontal=a to b]
    \begin{feynman}
      \vertex (a);
      \vertex [right=of a] (b);
      \vertex [right=of b] (c);
      \vertex [above left=of b] (d);
      \vertex [above right=of b] (e);
      \diagram{
        (a) -- [charged scalar, edge label'=\(p\)] (b) 
        -- [charged scalar, edge label'=\(p\)] (c),
        (d) -- [momentum={[arrowlabel]$k$}] (b),
        (b) -- [momentum={[arrowlabel]$k$}] (e),
      };
    \end{feynman}
  \end{tikzpicture}
  \nonumber \\
  &=& \quad
  F_{\textbf{IV}}(p^2,\bar{A},a) 
   \nonumber \\
   &=& \quad 
   \frac{g^{2} N}{V(N^{2}-1)}
   \frac{1}{p^2}
   \int_{k^{d}} \,
   \bar{A}^{d}_{\mu}(k)  \bar{A}^{d}_{\mu}(-k)
\end{eqnarray}

\begin{eqnarray}
   \text{(V)} 
\quad &=& \quad
     \begin{tikzpicture}[
  arrowlabel/.style={
    /tikzfeynman/momentum/.cd, 
    arrow shorten=#1,arrow distance=1.5mm},
    arrowlabel/.default=0.4,
  baseline=(b.base),horizontal=a to b]
    \begin{feynman}
      \vertex (a);
      \vertex [right=of a] (b);
      \vertex [right=of b] (c);
      \vertex [above left=of b] (d);
      \vertex [above right=of b] (e);
      \diagram{
        (a) -- [charged scalar, edge label'=\(p\)] (b) 
        -- [charged scalar, edge label'=\(p\)] (c),
        (d) -- [momentum={[arrowlabel]$k$}] (b),
        (b) -- [photon, momentum={[arrowlabel]$k$}] (e),
      };
    \end{feynman}
  \end{tikzpicture}
  \nonumber \\
  &=& \quad
  F_{\textbf{V}}(p^2,\bar{A},a)
   \nonumber \\
   &=& \quad 
   \frac{g^{2} N}{V(N^{2}-1)}
   \frac{1}{p^2}
   \int_{k^{d}}
   \bar{A}^{d}_{\mu}(k)  a^{d}_{\mu}(-k)
\end{eqnarray}  
  
\begin{eqnarray}
   \text{(VI)} 
\quad &=& \quad
     \begin{tikzpicture}[
  arrowlabel/.style={
    /tikzfeynman/momentum/.cd, 
    arrow shorten=#1,arrow distance=1.5mm},
    arrowlabel/.default=0.4,
  baseline=(b.base),horizontal=a to b]
    \begin{feynman}
      \vertex (a);
      \vertex [right=of a] (b);
      \vertex [right=of b] (c);
      \vertex [right=of c] (d);
      \vertex [above=of b] (e);
      \vertex [above=of c] (f);
      \diagram{
        (a) -- [charged scalar, edge label'={$p$}] (b) 
        -- [charged scalar, edge label'={$p+k$}] (c)
        -- [charged scalar, edge label'={$p$}] (d),
        (e) -- [photon, momentum={[arrowlabel]$k$}] (b),
        (c) -- [photon, momentum={[arrowlabel]$k$}] (f),
      };
    \end{feynman}
  \end{tikzpicture}
  \nonumber \\
  &=& \quad
  \sigma_{\textbf{VI}}(p^2,\bar{A},a)
   \nonumber \\
   &=& \quad 
   \frac{g^2 N}{V(N^{2}-1)} 
   \frac{p_{\mu}p_{\nu}}{p^{2}}
   \int_{k^{d}}
   \frac{ a^{d}_{\mu}(k) a^{d}_{\nu}(-k) }{(p+k)^{2}}
\end{eqnarray}

\begin{eqnarray}
   \text{(VII)} 
\quad &=& \quad
     \begin{tikzpicture}[
  arrowlabel/.style={
    /tikzfeynman/momentum/.cd, 
    arrow shorten=#1,arrow distance=1.5mm},
    arrowlabel/.default=0.4,
  baseline=(b.base),horizontal=a to b]
    \begin{feynman}
      \vertex (a);
      \vertex [right=of a] (b);
      \vertex [right=of b] (c);
      \vertex [right=of c] (d);
      \vertex [above=of b] (e);
      \vertex [above=of c] (f);
      \diagram{
        (a) -- [charged scalar, edge label'={$p$}] (b) 
        -- [charged scalar, edge label'={$p+k$}] (c)
        -- [charged scalar, edge label'={$p$}] (d),
        (e) -- [momentum={[arrowlabel]$k$}] (b),
        (c) -- [momentum={[arrowlabel]$k$}] (f),
      };
    \end{feynman}
  \end{tikzpicture}
  \nonumber \\
  &=&\quad
  \sigma_{\textbf{VII}}(p^2,\bar{A},a)
   \nonumber \\
   &=&\quad 
   \frac{4g^{2}N}{V(N^{2}-1)}
   \int_{k^{d}}
   \frac{ \bar{A}^{d}_{\mu}(k)   \bar{A}^{d}_{\nu}(-k) }{(p+k)^{2}}
\end{eqnarray}

\begin{eqnarray}
   \text{(VIII)} 
\quad &=& \quad
     \begin{tikzpicture}[
  arrowlabel/.style={
    /tikzfeynman/momentum/.cd, 
    arrow shorten=#1,arrow distance=1.5mm},
    arrowlabel/.default=0.4,
  baseline=(b.base),horizontal=a to b]
    \begin{feynman}
      \vertex (a);
      \vertex [right=of a] (b);
      \vertex [right=of b] (c);
      \vertex [right=of c] (d);
      \vertex [above=of b] (e);
      \vertex [above=of c] (f);
      \diagram{
        (a) -- [charged scalar, edge label'={$p$}] (b) 
        -- [charged scalar, edge label'={$p+k$}] (c)
        -- [charged scalar, edge label'={$p$}] (d),
        (e) -- [momentum={[arrowlabel]$k$}] (b),
        (c) -- [photon, momentum={[arrowlabel]$k$}] (f),
      };
    \end{feynman}
  \end{tikzpicture}
  \nonumber \\
  &=&\quad
  \sigma_{\textbf{VIII}}(p^2,\bar{A},a)
   \nonumber \\
   &=&\quad 
   \frac{2g^{2}N}{4V(N^{2}-1)}
   \int_{k^{d}}  \;
   \frac{ \bar{A}_{\mu}(k) a_{\mu}(-k)  }{(p+k)^{2}}
\end{eqnarray}

\bibliography{refs}

\end{document}